\documentclass[aps,prb,twocolumn]{revtex4}

\usepackage{graphicx}

\begin{document}

\title{Phase dependence of localization in the driven two-level model}

\author{C.E.~Creffield}
\affiliation{Dipartimento di Fisica, Universit\`a di Roma ``La Sapienza'',
Piazzale Aldo Moro 2, I-00185, Rome, Italy}

\date{\today}


\begin{abstract}
A two-level system subjected to a high-frequency
driving field can exhibit an effect termed ``coherent destruction
of tunneling'', in which the tunneling of the system is suppressed
at certain values of the frequency and strength
of the field. This suppression becomes less
effective as the frequency of the driving field is reduced, and
we show here how the detailed form of its fall-off depends
on the {\em phase} of the driving, which for certain values
can produce small local maxima (or revivals) in the overall decay.
By considering a squarewave driving field,
which has the advantage of being analytically tractable, we show
how this surprising behavior can be interpreted geometrically in terms of
orbits on the Bloch sphere. These results are of general applicability
to more commonly used fields, such as sinusoidal driving,
which display a similar phenomenology.
\end{abstract}


\maketitle

Since the pioneering work of H\"anggi and co-workers
\cite{hanggi_prl} it has been known that a quantum
system subjected to a periodic driving field can exhibit
an unusual effect termed ``coherent destruction of tunneling'' (CDT),
in which the tunneling dynamics of the system are quenched at various
specific values of the strength and frequency 
\cite{holthaus} of the applied field.
One of the simplest systems in which this effect appears is the driven
two-level system \cite{hanggi_epl}, which has been applied to describe many
physical situations, such as an electron moving in a double-well
potential \cite{hanggi_prl}, molecular switches \cite{lehmann},
the single-Cooper-pair box \cite{nakamura}
and superconducting tunnel junctions \cite{vion}.
Controlling the dynamics of such systems has become of increasing importance
due to possible technological applications in quantum computing, and the
CDT effect provides one method of achieving such control which preserves the
quantum coherence of the system.

The periodicity of the driving field allows us to invoke the Floquet
theorem, and expand solutions of the time-dependent Schr\"odinger
equation \cite{review} as
\begin{equation}
\psi(t) = \sum_{j=1}^2 \ c_j \exp[-i \epsilon_j t] \phi_j(t) ,
\label{expand}
\end{equation}
where $\epsilon_j$ is termed the quasienergy, and $\phi_j(t)$ is a function
with the same periodicity as the driving field, called a Floquet state.
A {\em necessary} condition for CDT to occur is that the
quasienergies must be either degenerate  or close to degeneracy \cite{lehmann}.
If we neglect the time-dependence of the Floquet states it is clear
that the dynamics of the system will indeed be frozen when this
condition is satisfied.
Neglecting the intrinsic time-dependence of the Floquet states
is a reasonable approximation when the frequency of the driving field is
high (the ``high-frequency limit'' \cite{hanggi_epl}),
as the period of the field is then much shorter than the typical timescale
for tunneling processes, given by the Rabi period. At lower frequencies,
however, the Floquet states {\em can} have a non-trivial time-dependence,
and the degree of CDT can 
be markedly reduced if the Floquet states themselves have a large amplitude
of oscillation \cite{caveat}. In this work we will show that 
fall-off in CDT produced by this effect
has a surprising dependence on the {\em phase} of the driving field.
Although phase-effects have been briefly noted in previous work
\cite{bavli}, we provide here the first systematic study and explanation of
this effect. We shall show how the phase can be chosen to maximize
the degree of CDT, and also how certain choices of phase can
produce a non-monotonic decay of CDT, containing small local maxima
or ``revivals'', as was seen previously in Ref.\cite{cabron}.

We consider a two-level system described by the Hamiltonian 
\begin{equation}
H = \frac{\Delta}{2} \ \sigma_x + \frac{E}{2} f(t) \ \sigma_z ,
\label{hamilton}
\end{equation}
where $\sigma_i$ are the standard Pauli matrices.
We parameterize the driving field as $E f(t)$, where 
$E$ is the strength of the driving, and $f(t)$ is
a $T$-periodic function of unit amplitude and frequency $\omega$,
which describes the waveform of the driving field.
The eigenstates of the undriven system
consist of two {\em extended} states: namely a symmetric ground state
separated by the level splitting, $\Delta$, from the anti-symmetric
excited state. To investigate the tunneling dynamics of the system
it will prove useful to also define {\em localized} states formed
by the sum and difference of these eigenstates, which we label
as $| L \rangle$ and $| R \rangle$.
The degree of tunneling can be conveniently assessed by initializing
the system in state $| L \rangle$ and then evolving it in time under the
influence of the Hamiltonian (\ref{hamilton}). By measuring the probability
that the system remains in state $| L \rangle$,
$P_L (t) = \left| \langle \psi(t) | L \rangle \right|^2$,
we can quantify the degree to which tunneling is suppressed by finding
the minimum value of $P_L(t)$ attained, which we term the ``localization''.
When the tunneling is completely destroyed $P_L(t)$
does not change with time, and thus the localization
is equal to one. Conversely, if the tunneling is not completely suppressed,
the localization will take a lower value, and will be zero if the particle
is able to completely tunnel from $| L \rangle$ to $| R \rangle$.

\begin{figure}
\centerline{\includegraphics[width=0.35\textwidth,clip]{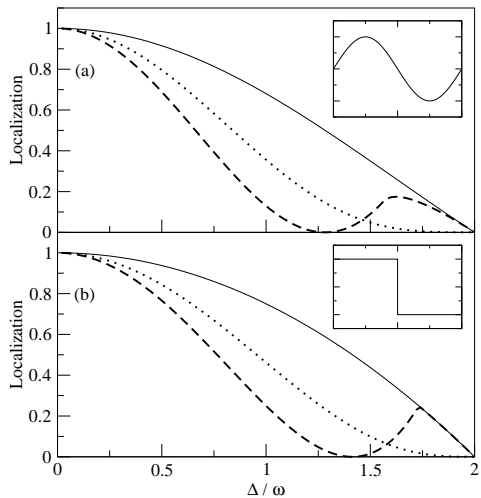}}
\caption{Localization produced by (a) a sinusoidal driving field
and (b) a squarewave driving field. 
Solid line indicates the result for a driving with
zero phase-shift, the dotted line for
a phase-shift of $\pi / 4$, and the dashed line for a phase-shift of $\pi/2$.
For high-values of $\omega$
the localization is excellent in all cases, but drops rapidly for
$\Delta / \omega > 0.5$. For the phase-shift of $\pi/2$ the localization
goes through a small local minimum
before reaching zero. Insets: A single period of the zero phase-shifted
driving fields $f(t)$.}
\label{local}
\end{figure}

We show in Fig.\ref{local} the localization produced
as the frequency of the driving field is reduced, obtained 
by the numerical evolution of the system using a
Runge-Kutta technique \cite{creff},
for two types of driving field: a sine wave and a squarewave. 
For both fields the driving strength
was chosen so that the quasienergies of the 
system were degenerate and so CDT occurs. As was shown in Ref.\cite{creff},
this degeneracy condition defines a set of smooth
one-dimensional curves in $(E, \omega)$ space called 
``crossing-manifolds'' \cite{hanggi_epl}. 
When the quasienergies are degenerate it is clear from
Eq.(\ref{expand}) that the time-dependence of
the system is given {\em solely} by the Floquet states. Since these states
are periodic, it is sufficient to evolve the system over
{\em just one period} of the driving to study its properties,
giving a considerable saving of computational effort.

Although the two driving fields are very different,
they produce strikingly similar forms of localization.
In both cases
in the high-frequency regime ($\Delta / \omega \ll 1$)
the localization has a value of one, irrespective
of the phase of the driving field. This indicates that the tunneling
is completely destroyed and thus the system remains frozen in its
initial state. As the value of $\omega$ is reduced, 
the particle is able to tunnel more freely, 
and the degree of localization falls.
It can also be clearly seen that at low frequencies
the phase of the driving field has a significant effect on the 
form of this fall-off. When the field has {\em no} phase-shift 
(shown in the insets to Fig.\ref{local})
the localization decays monotonically to zero
as the frequency is reduced.
The initial effect of increasing the phase of the driving field
from zero
is to cause the localization to drop more rapidly. For example,
at $\Delta / \omega = 1$ the localization produced by a $\pi / 4$
phase-shifted field is half that of the zero phase-shift case,
while for a phase shift of $\pi / 2$ the localization is
about four times smaller. A further effect occurs for values of
the phase shift close to its maximum value of $\pi / 2$, in which 
after rapidly decaying to zero,the localization subsequently
passes through a small local maximum, before again 
vanishing at the same frequency as the zero phase-shift case.

To understand how this behavior comes about, it is
useful to visualize the dynamics of the two-level system
geometrically, by making use of the Bloch sphere representation
well-known from quantum optics and NMR studies, and now
commonly used in quantum information theory \cite{schirmer}.
Any pure state of the system can be represented by a
point on the surface of the Bloch sphere, 
and we can note in particular that the localized states
$| L \rangle$ and $| R \rangle$ correspond to the
points $[0,0,\pm 1 ]$.
We shall consider just the case of squarewave driving, as
it displays the same phase-dependent behavior as sinusoidal
driving, but also has the useful property of allowing
exact solutions for the time-development
of the system to be obtained. For this form
of driving, the time-dependent
Hamiltonian (\ref{hamilton}) decomposes into 
piecewise constant parts, $H_{\pm} = ( \Delta / 2 ) \sigma_x
\pm (E / 2) \sigma_z$. These 
operators can be interpreted conveniently
as an interaction between the Bloch vector ${\bf s}$ and a
fictitious magnetic field directed
along the axes $r_{\pm}=[\Delta,0,\pm E]$, and thus the
time-evolution of the system can be viewed as successive Larmor
rotations about these axes. This insight is
the key to a simple method for interpreting and understanding
the dynamics of the system.

The operators $H_{\pm}$ can be exponentiated
straightforwardly, yielding the following result
for the unitary operator describing the evolution of the
system under the influence of $H_{\pm}$:
\begin{equation}
U_{\pm}(t) = I \cos \Omega t - i \frac{\sin \Omega t}{2 \Omega}
\left[ \Delta \ \sigma_x \pm E \ \sigma_z \right] .
\label{unitary}
\end{equation} 
In this expression $\Omega=\sqrt{E^2 + \Delta^2} / 2$ is the Rabi 
frequency. The general time-evolution operator 
for a squarewave driving field
can now be expressed simply as the product of
successive factors of $U_{\pm}$. Examination of Eq.\ref{unitary}
reveals that the propagator for one period of the driving,
$U(T) = U_{+}(T/2) U_{-}(T/2)$, is equal to the identity
(and thus the quasienergies are both zero and CDT occurs) 
if $\sin \Omega T/2 = 0$.
This indicates that CDT only arises if the driving
frequency is in resonance with the Rabi frequency,
$\Omega = n \omega$ where $n$ is an integer. This resonance
condition may be expressed alternatively as
\begin{equation}
\left( \frac{E}{\omega} \right)^2 +
\left( \frac{\Delta}{\omega} \right)^2 = \left( 2 n \right)^2 .
\label{manifold}
\end{equation}
It is interesting to note that this equation for the crossing-manifolds
for a squarewave driving field,
which describes concentric circles in $(E/\omega,\Delta/\omega)$ space, 
is identical 
to that hypothesized from purely numerical evidence in Ref.\cite{creff}. 

\begin{figure}
\centerline{\includegraphics[width=0.4\textwidth,clip]{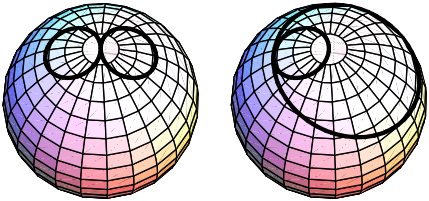}}
\caption{Orbits on the Bloch sphere for driving parameters
$\Delta/\omega=1/2$, $E/\omega=\sqrt{15}/2$,
{\it i.e.} on the first crossing-manifold ($n=1$).
The left plot shows the time-evolution for zero phase-shift, 
and on the right for a phase-shift of $\pi/2$.}
\label{sphere}
\end{figure}

\begin{figure}
\centerline{\includegraphics[width=0.35\textwidth,clip]{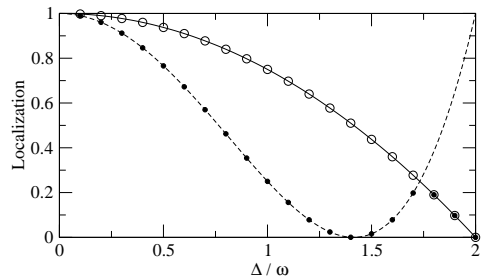}}
\caption{Comparison of the localization  
on the first crossing-manifold ($n=1$) with the analytical
expressions derived in the text.
Eq.\ref{exact_1} is denoted by the solid line,
Eq.\ref{exact_2} by the dashed line.
It can be seen that for zero-phase shifted 
driving (hollow circles) the localization
is described by Eq.\ref{exact_1}.
For $\pi/2$ phase-shifted driving (solid circles) 
the localization is given by min $[P_L(T/4), P_L(T/2)]$,
and the crossover between these quantities at 
$\Delta / \omega = \sqrt{3}$ produces the cusp-like revival feature.}
\label{compare}
\end{figure}

We now consider the explicit time evolution of the system, and
begin with the simplest case of zero phase-shift. 
The state $| L \rangle$ in which the system is initialized
corresponds to the north-pole of the Bloch sphere (${\bf s}=[0,0,1]$). 
During the first half-period of the driving, this initial state
evolves under the influence of $H_+$, making a Larmor rotation about
the axis $r_{+}$. The resonance condition 
between the driving frequency $\omega$ and the Larmor frequency
means that during this interval the Bloch vector makes $n$ complete
revolutions, and thus at $t=T/2$ the system returns
to its initial state. During the second half-period a similar
process occurs, with the Bloch vector making $n$ revolutions about the
axis $r_{-}$. The system thus traces out a symmetric
``figure-of-eight''
on the surface of the Bloch sphere, as shown in Fig.\ref{sphere}.
In the high-frequency limit, when $\omega \gg \Delta$
and thus $E \sim \omega$ from Eq.\ref{manifold}, the rotation axes
are almost parallel to the $z$-axis, and so the loops
traced by the Bloch vector have a very small radius.
As a result, the orbit remains close to $| L \rangle$ at all times, 
and the system exhibits a high degree of 
localization. As $E$ is reduced, however, the angle between the rotation
axes increases and the loops traced by the Bloch vector become larger.
The system can thus explore further around the Bloch
sphere and approach closer to $| R \rangle$, 
thereby reducing the localization. 
It is a simple matter to explicitly derive an expression for
$P_L(t)$, from which the following result
for the localization can be obtained
\begin{equation}
P_L^{\mbox{min}} = P_L(T/4) = 
1 - \frac{ \left( \Delta/\omega \right)^2 }{4 n^2} .
\label{exact_1}
\end{equation}
In Fig.\ref{compare} it can be seen that this result indeed 
correctly describes
the fall-off in localization for a zero phase-shifted field.

We now consider the effect of the phase-shift.
For a phase-shift of $\pi/2$ the system first evolves
under $H_{+}$ for a quarter-period, then under $H_{-}$ for an
interval of $T/2$, and completes the final quarter-period under $H_{+}$.
Accordingly, during the first quarter-period of the driving
the Bloch vector makes $n/2$ revolutions about $r_+$, and
it is thus now important to
distinguish whether $n$ is even or odd.
In the former case, the Bloch vector will 
make an integer number of revolutions about $r_{+}$
in the first quarter-period, and so it will end this
interval at $| L \rangle$. It will then make $n$ complete revolutions
about $r_{-}$ during the next half-period,
and will finally complete its motion by making $n/2$ revolutions
about $r_{+}$. For even values of $n$ the Bloch vector thus
traces out the same figure-of-eight orbit as for
the zero phase-shift case, and so although
the order in which it traverses the loops is different,
identical values of localization will be produced.

If $n$ is odd, however, a different behavior will occur.
At the end of the first quarter-period the system
will {\em not} have returned to $| L \rangle$, as the Bloch vector
will only have made a {\em half-integer} number of rotations about $r_+$.
The Bloch vector has thus been rotated away
from the $r_{-}$ axis, and so under the influence of $H_{-}$
the Larmor rotation it makes around this axis will have a 
different radius, as shown in Fig.\ref{sphere}.
The second loop clearly
approaches much closer to $| R \rangle$ than for
the case of the figure-of-eight orbit,
and thus the localization is reduced. 

\begin{figure}
\centerline{\includegraphics[width=0.45\textwidth,clip]{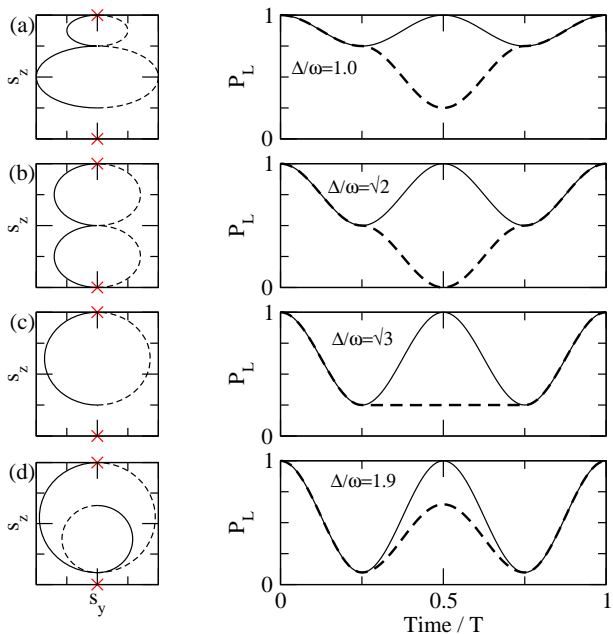}}
\caption{Time-evolution of the system in
the low-frequency regime $\omega \leq \Delta$.
On the left, the evolution on the Bloch
sphere projected onto the $(s_y,s_z)$ plane. 
The solid line indicates the evolution of the system
over the first half-period, the dotted line
over the second half-period.
The upper (lower) crosses indicate the states $| L \rangle$
($| R \rangle$). On the right, the time-evolution of the 
probability $P_L(t)$ over one period of the driving. The solid line
indicates zero phase-shifted driving, the dashed line has
a phase-shift of $\pi/2$.}
\label{orbits}
\end{figure}

In Fig.\ref{orbits} we show this effect in detail 
for driving parameters on the first-crossing manifold ($n=1$),
by displaying the Bloch sphere orbits projected onto
the $(s_y, s_z)$ plane. For $\Delta/\omega=1$ the system makes
a similar orbit to that shown in Fig.\ref{sphere}.
The minimum value of localization 
({\it i.e.} the closest approach to $| R \rangle$) occurs at $t=T/2$, 
in contrast to the case of zero-phase shift, for which the minima 
occur at $t=T/4$ and $3 T/4$. Although it is possible to calculate 
explicit expressions for $P_L(t)$, these are in general somewhat 
cumbersome. We instead present formulae just for times that are
multiples of $T/4$, which, as can be seen from Fig.\ref{orbits},
correspond to the turning points in $P_L(t)$.

During the first and last quarter-periods of driving
the time-evolution of the system is identical
to that of the case with zero phase-shift, and accordingly
$P_L(T/4)$ is given by Eq.\ref{exact_1}. It is also straightforward
to show that
\begin{equation}
P_L(T/2) = \left[ 1 - \frac{ \left( \Delta / \omega \right)^2}{2 n^2}
\sin^2 \pi n /2 \right]^2 .
\label{exact_2}
\end{equation}
In Fig.\ref{compare} it can be seen that this expression indeed
correctly describes the rapid fall of localization
for the $\pi/2$ phase-shifted field, which
reaches zero at $\Delta / \omega = \sqrt{2}$.
At this value of frequency,
the two rotation axes make an angle of $\pi/4$ to the $z$-axis,
producing the highly symmetric time-evolution shown
in Fig.\ref{orbits}b, in which the system
passes through $| R \rangle$ at $t=T/2$.
As $\Delta / \omega$ is increased further, however, the trajectory
of the orbit no longer passes exactly through $| R \rangle$ and thus
the localization increases, producing the ``revival'' effect.
This increase continues until $\Delta / \omega = \sqrt{3}$,
at which $P_L (T/4)$ and $P_L(T/2)$ become equal. As shown in
Fig.\ref{orbits}c, at this value of frequency the system does not
display any time evolution during the interval $(T/4,3T/4)$, 
since the action of first quarter-period of the driving is to rotate the
initial state to align with $r_{-}$. The following half-period of driving
under $H_{-}$, thus does not alter the state of the system.

For still larger values of $\Delta / \omega$, the rotation
about the $r_{-}$ axis acts to move the system {\em away} from
$| R \rangle$ (see Fig.\ref{orbits}d), and consequently the minimum
value of $P_L$ no longer occurs at $T/2$, but at $T/4$ instead. 
In this regime the localization produced by the 
$\pi/2$ phase-shifted field is therefore identical to that
produced by the zero phase-shifted case.
It can be clearly seen in Fig.\ref{compare} that 
it is this crossover in behavior
that gives rise to the revival feature seen
in Fig.\ref{local}.  
  
In summary, we have studied how the CDT effect depends upon the 
frequency of the driving field. 
In the high-frequency regime, the positions of the quasienergy
degeneracy can be found analytically \cite{holthaus,creff}
and the degree of CDT is not affected by the phase of the driving field.
As the frequency is reduced a more complicated picture appears:
the quasienergies drift away from their high-frequency values
along one-dimensional manifolds, and the degree of CDT induced depends not
only on the values of field strength and frequency, but also on its phase.
We have firstly shown that the form of the crossing-manifolds
for the squarewave driving field
(Eq.\ref{manifold}) that was conjectured previously on numerical evidence 
\cite{creff} is in fact {\em exact}. By using the Bloch sphere
representation we have given a simple method of understanding 
how the localization falls as the frequency of the driving is reduced,
and have analytically confirmed the empirical
observation in Ref.\cite{bavli} that the
localization is maximized for zero phase-shifted driving.
We have also clarified how the revival features seen for $\pi/2$
phase-shifted driving \cite{cabron} arise, and have demonstrated
that for crossing-manifolds with even values of $n$ that
this phenomenon does not occur.
This opens up new prospects for experiment in the low-frequency
regime, which is both easier to attain and also
has the advantage that the driving field is less likely
to drive transitions to higher energy-states (thereby breaking the 
two-level approximation).

\acknowledgments
The author gratefully acknowledges the hospitality of the 
University of Edinburgh, where this work was completed.


\begin{thebibliography}{}
\bibitem{caveat}
{In this work we monitor the tunneling dynamics of the system
continuously. If, however, the system is measured ``stroboscopically''
at intervals separated by the period of the driving field, CDT will
always appear to be perfect if the condition of quasienergy degeneracy
is satisfied.}
\end{thebibliography}
\end{document}